# The role of electron capture and energy exchange of positively charged particles passing through matter

W. Ulmer

Klinikum München-Pasing, Dept. of Radiotherapy and MPI of Physics, Göttingen, Germany

e-mail: waldemar.ulmer@gmx.net

**Abstract**

The conventional treatment of the Bethe-Bloch equation for protons accounts for electron capture at the end of the projectile track by the small Barkas correction. This is only a possible way for protons, whereas for light and heavier charged nuclei the exchange of energy and charge along the track has to be accounted for by regarding the projectile charge q as a function of the residual energy. This leads to a significant modification of the Bethe-Bloch equation, otherwise the range in a medium is incorrectly determined. The LET in the Bragg peak domain and distal end is significantly influenced by the electron capture. A rather significant result is that in the domain of the Bragg peak the superiority of carbon ions is reduced compared to protons.

1. **Introduction**

The application of the Bethe-Bloch equation (BBE) for the determination of the electronic stopping power is established for the passage of electrons and protons through homogeneous media. A particular importance of BBE appears in Monte-Carlo calculations to simulate behavior of charged projectile particles along the track (e.g. the code GEANT4). This equation reads:

$$-dE(z)/dz = (K/v^2) \cdot [\ln(2mv^2/E_I) - \ln(1-\beta^2) + \\ + a_{shell} + a_{Barkas} + a_0 v^2 + a_{Bloch}] \\ K = (Z\rho/A_N) \cdot 8\pi q^2 e_0^4 / 2m \qquad (1)$$

$E_I$ is the atomic ionization energy, weighted over all possible transition probabilities of atomic/molecular shells, and q denotes the charge number of the projectile (e.g. proton). The meaning of the correction terms $a_{shell}$, $a_{Barkas}$, $a_0$ and $a_{Bloch}$ are explained in literature, see Bethe (1930), Bloch (1933), Bethe et al (1953), Bethe (1953), ICRU49 (1993), and Boon (1998). Since the Bloch correction $a_{Bloch}$ will be introduced in equation (12), we present, for completeness, the remaining correction terms according to ICRU49:

$$a_0 : \text{constant value}; \; ICRU\, 49: a_0 = -1 \qquad (2)$$

$$a_{shell} = -C/Z \qquad (3)$$

$$a_{Barkas} = \sqrt{2} \cdot F_{ARB}(b/\sqrt{\kappa})/(\sqrt{Z} \cdot \kappa^{3/2}) \\ \kappa = Z^{-1} \cdot (\beta/\alpha)^2 \qquad (4)$$

Some comments to the relations (2 – 4):

The parameter C, referring to shell corrections, is determined by different models (ICRU49 and references therein). A unique parameterization of C depending on Z, $A_N$, and $E_I$ does not exist. It is therefore recommended to select C according to proper domains of validity. It must be noted that several models have been proposed to account for shell transitions. Therefore, the recommendations of ICRU49 have been applied in this work.

The function $F_{ARB}$ in equation (4) refers to the theory of the Barkas effect developed by Ashley, Ritchie and Brandt (Ashley et al 1974). The parameter α refers to Sommerfeld's fine structure constant and b to a fitting parameter. Unfortunately, b is not a unique fitting parameter; this results in an uncertainty of about 2 %. Modifications of equation for high-Z materials are not of interest in this work. The Barkas effect represents a correction of BBE due to the electron capture of the positively charged protons at lower energies in the domain of the Bragg peak and behind leading to a slightly increased range $R_{csda}$, whereas the negatively charged anti-protons cannot capture electrons from the environmental electrons. Therefore their range is slightly smaller. With regard to protons this kind of correction works, i.e. the charge $q^2 = 1$ is assumed along the total proton track, whereas for charged ions such as He or $C^6$ is appears to be insufficient to keep the nuclear charge constant along the total track and to restrict the electron capture only to the small Barkas correction (Barkas et al 1963). This means that all positively charged projectile particles stand in permanent exchange of energy E and charge q with environment, and, as a consequence, $q^2$ is a function of the actual residual energy, i.e. $q^2 = q^2(E)$, and only for $E = E_0$ (initial energy) $q^2 = q_0^2$ is valid. A correct modification of BBE by accounting for $q^2(E)$ makes the Barkas correction superfluous.

A further critical aspect of BBE, which leads to a modification by accounting for $q^2(E)$ is the range $R_{csda}$ of the electronic stopping power. Thus a naïve application of BBE would lead to the conclusion that a carbon ion would require the initial energy per nucleon $E_0$ (carbon ion) = 3 x $E_0$(proton), since the square of the carbon charge amounts to 36 and the nuclear mass unit is 12 x nuclear mass unit of the proton. However, the ratio is not 3 to obtain the same range $R_{csda}$, but about 25/12. The Monte-Carlo code GEANT4 assumes an average charge $q_{Average}$ = 5.06 for the simulations of the carbon tracks. This is, however, not satisfactory, since electron capture is a dynamical process. Therefore the range of carbon ions has been subjected to many studies (Betz 1972, Dingfelder et al 1998, Gudowska et al 2004, Hollmark et al 2004, Hubert et al 1990, Kanai et al 1993, Kusano et al 2007, Martini 2007, Matveev et al 2006, Sigmund et al 2006, Sigmund 2006, Sihver et al 1998, Yarlagadda et al 1978, Zhang and Newhauser 2009, Ziegler et al 1988) due to the increasing importance of carbon ions in radiotherapy.

It is also possible to substitute the electron mass m by the reduced mass m $\Rightarrow$ µ. However, this leads for protons to a rather small correction (i.e., less than 0.1 % for protons). For complex systems $E_I$ and some other contributions like $a_{shell}$ and $a_{Barkas}$ can only be approximately calculated by simple quantum-mechanical models (e.g., harmonic oscillator); the latter terms are often omitted and $E_I$ is treated as a fitting parameter, but different values are proposed and used (ICRU49). The restriction to the logarithmic term leads to severe problems, if either $v \to 0$ or $2m v^2 /E_I \to 1$. It should be added that a correct treatment of the electron capture removes the singularity of positively charged ions, since $q^2(E) \to 0$, if the residual energy E assumes zero.

## 2. Methods

**2.1 The integration BBE for protons**

In the following, we consider at first the integration of BBE for protons, i.e. we consider the Barkas correction in the conventional way. In previous publications (Ulmer 2007, Ulmer and Matsinos 2010) we have presented an analytical integration of BBE. BBE is the physical base in the transport of protons and electrons.

In order to obtain the integration of BBE, we start with the logarithmic term and perform the substitutions:

$$v^2 = 2E/M; \quad \beta_I = 4m/ME_I; \quad E = (1/\beta_I)\exp(-u/2) \quad (5)$$

With the help of substitution (and without any correction terms), BBE leads to the integration:

$$\left. \begin{array}{l} -\int du \exp(-u) \cdot (1/u) = \tfrac{1}{2} K \cdot \beta_I^2 \cdot M \int dz \\ K = (Z\rho/A_N) \cdot 8\pi q^2 e_0^4 / 2m \end{array} \right\} \quad (6)$$

The boundary conditions of the integral are:

$$\left. \begin{array}{l} z = 0 \Rightarrow E = E_0 \text{ (or : } u = -2\ln(E_0 \cdot \beta_I)) \\ z = R_{CSDA} \Rightarrow E = 0 \text{ (or : } u \Rightarrow \infty) \end{array} \right\} \quad (7)$$

The general solution is given by the Euler exponential integral function Ei($\xi$) with P.V. = principal value:

$$\left. \begin{array}{l} \tfrac{1}{2} K \cdot M \cdot \beta_I^2 \cdot R_{CSDA} = -P.V. \int_{-\xi}^{\infty} u^{-1} \exp(-u)\, du = Ei(\xi) \\ \xi = 2 \ln(4mE_0 / ME_I) \quad \text{and} \quad \xi > 0 \end{array} \right\} \qquad (8)$$

Some details of $Ei(\xi)$ and its power expansions can be found in Abramowitz and Stegun (1970). The critical case $\xi = 0$ results from $E_{critical} = ME_I/4m$ (for water with $E_I = 75.1$ eV, the critical energy $E_{critical}$ amounts to 34.474 keV; for Pb with $E_I \approx 800$ eV to about 0.4 MeV). Since the logarithmic term derived by Bethe implies the Born approximation, valid only if the transferred energy $E_{transfer} \gg$ the energy of shell transitions, the above corrections, exempting the Bloch correction, play a significant role in the environment of the Bragg peak, and the terms $a_0$ and $a_{shell}$ remove the singularity. With respect to numerical integrations (Monte Carlo), we note that, in the environment of $E = E_{critical}$, the logarithmic term may become crucial (leading to overflows); rigorous cutoffs circumvent the problem. Therefore, the shell correction is an important feature for low proton energies. In similar fashion, we can take account of the Barkas correction. Since this correction is also important for low proton energies, it is difficult to make a quantitative distinction to the shell correction, and different models exist in the literature implying overall errors up to 2 % (ICRU49). Using the definitions/suggestions of the correction terms according to ICRU49 and the substitutions we obtain:

$$\left. \begin{array}{l} \tfrac{1}{2} K \cdot M \cdot \beta_I^2 \int dz = \int du \exp(-u)[u + 2\alpha_S + \\ + 2\alpha_{Barkas}(4 \cdot m / E_I)^{p_B} \exp(p_B u / 2) + \alpha_0(E_I / 2m)\exp(-u/2)]^{-1} \end{array} \right\} \qquad (9)$$

A closed integration of equation (9) does not exist, but it can be evaluated via a procedure valid for integral operators (Feynman 1962), which reads for commutative operators:

$$[A' + B']^{-1} = A'^{-1} - A'^{-2} B' + A'^{-3} B'^2 - A'^{-4} B'^3 + \ldots + (-1)^n A'^{-n-1} B'^n \qquad (10)$$

A'+B' is equated to the complete denominator on the right-hand side of equation . The small Barkas correction and the Bloch correction $a_{Bloch}$ (see equations 12,13) are identified with A' and the other (more important) terms with B':

$$\left. \begin{array}{l} A' = 2\alpha_{Barkas}(4m / E_I)^{p_B} \exp(p_B u / 2) + a_{Bloch} \\ B' = u + 2\alpha_S + \alpha_0(E_I / 2m)\exp(-u/2) \end{array} \right\} \qquad (11)$$

We should already point out here that the expansion (10) is also applicable, if the electron capture is accounted for. The integration of equation with the help of relation (10) leads to standard tasks (i.e. to a series of usual exponential functions). In the following, we add the Bloch correction to the denominator of equation. In order to use the procedure, we define now the non-relativistic energy $E_{nr}$ by: $E_{nr} = 0.5 \cdot Mv^2$ and write the relativistic energy expression $E_{rel}$ (the rest energy $Mc^2$ is omitted) in terms of an expansion:

$$\left. \begin{array}{l} a_{Bloch} = -(q^2\alpha^2/\gamma^2)[1.042 - 0.8549\, q^2\alpha^2/\gamma^2 + \\ 0.343\, q^4\alpha^4/\gamma^4 - . + \text{higher} \quad \text{order} \quad \text{terms}\,] \\ \alpha = 1/137.036\,, \\ \gamma^2 = 2E_{nr}/Mc^2 = 2\exp(-u/2)/\beta\, Mc^2 \end{array} \right\} \quad (12)$$

Relation (12) provides a sequence of exponential functions:

$$\left. \begin{array}{l} a_{Bloch} = -(\tfrac{1}{2}q^2\alpha^2\beta_I Mc^2 \exp(u/2))[1.042 - \tfrac{1}{2}0.854 \cdot q^2\alpha^2\beta_I Mc^2 \exp(u/2) + \\ +\tfrac{1}{4}0.343 q^4\alpha^4\beta_I^2 M^2 c^4 \exp(u) - .. + \text{higher} - \text{order} \quad \text{terms}] \end{array} \right\} \quad (13)$$

$$\tfrac{1}{2} K \cdot M \cdot \beta_I^2 \int dz = \int du\, \exp(-u) \cdot [A' + B']^{-1} \quad (14)$$

The integration of equation is carried out with the boundary conditions. Since these conditions are defined by logarithmic values, which have to be inserted to an exponential function series, the result yields a power expansion for $R_{CSDA}$ in terms of $E_0$:

$$R_{CSDA} = \frac{1}{\rho} \cdot \frac{A_N}{Z} \sum_{n=1}^{N} \alpha_n E_I^{pn} E_0^n \quad (N \Rightarrow \infty) \quad (15)$$

The coefficients $\alpha_n$ are determined by the integration procedure and only depend on the parameters of the BBE. For applications to therapeutic protons, i.e., $E_0 < 300$ MeV, a restriction to $N = 4$ provides excellent results (Figure 1). For water, we have to take $E_I = 75.1$ eV, $Z/A_N = 10/18$, $\rho = 1$ g/cm$^3$; Formula becomes:

$$R_{CSDA} = \sum_{n=1}^{N} a_n E_0^n \quad (N \Rightarrow \infty) \quad (16)$$

The values of the parameters of Formulas with restriction to N = 4 are displayed in Tables 1 and 2.

**Table 1:** Parameter values for equation (16) if $E_0$ is in MeV, $E_I$ in eV and $R_{CSDA}$ in cm.

| $\alpha_1$ | $\alpha_2$ | $\alpha_3$ | $\alpha_4$ | p1 | p2 | p3 | p4 |
|---|---|---|---|---|---|---|---|
| $6.8469 \cdot 10^{-4}$ | $2.26769 \cdot 10^{-4}$ | $-2.4610 \cdot 10^{-7}$ | $1.4275 \cdot 10^{-10}$ | 0.4002 | 0.1594 | 0.2326 | 0.3264 |

**Table 2:** Parameter values for equation (17), if $E_0$ is in MeV, $E_I$ in eV and $R_{CSDA}$ in cm.

| $a_1$ | $a_2$ | $a_3$ | $a_4$ |
|---|---|---|---|
| $6.94656 \cdot 10^{-3}$ | $8.13116 \cdot 10^{-4}$ | $-1.21068 \cdot 10^{-6}$ | $1.053 \cdot 10^{-9}$ |

The determination of $A_N$ and $Z$ is not a problem in case of atoms or molecules, where weight factors can be introduced according to the Bragg rule; for tissue heterogeneities, it is already a difficult task. Much more difficult is the accurate determination of $E_I$, which results from transition probabilities of all atomic/molecular states to the continuum (δ-electrons). Thus, according to the report ICRU49 of stopping powers of protons in different media, there are sometimes different values of $E_I$ proposed (e.g., for Pb: $E_I$ = 820 eV and $E_I$ = 779 eV). If we use the average (i.e., $E_I$ = 800.5 eV), the above formula provides a mean standard deviation of 0.27 % referred to stopping-power data in ICRU49, whereas for $E_I$ = 820 eV or $E_I$ = 779 eV we obtain 0.35 % - 0.4 %. If we apply the above formula to data of other elements listed in ICRU49, the mean standard deviations also amount to about 0.2 % - 0.4 %.

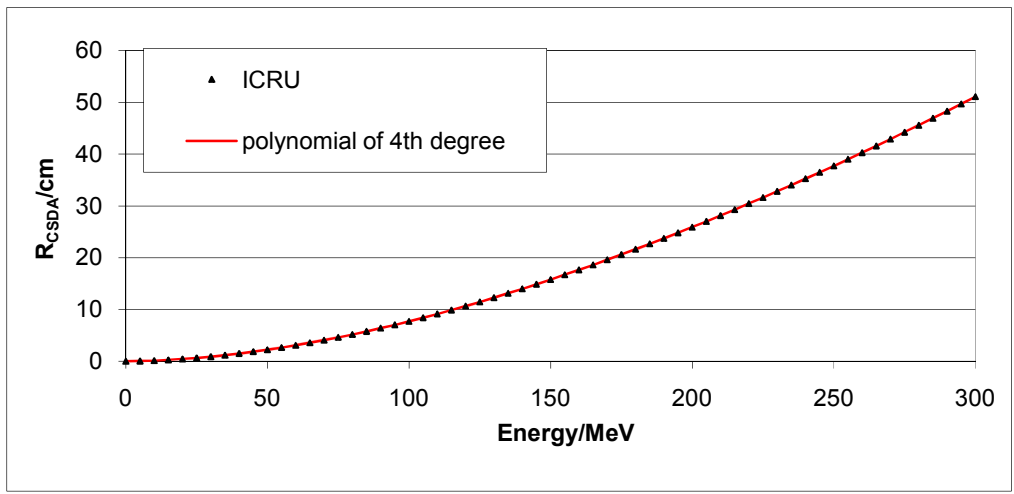

**Figure 1:** Comparison between ICRU49 data of proton $R_{CSDA}$ range (up to 300 MeV) in water and the fourth-degree polynomial (equation 16). The average deviation amounts to 0.0013 MeV.

Instead of the usual power expansion (16), we can represent all integrals in terms of Gompertz-type functions multiplied with a single exponential function by collection of all exponential functions obtained by the expression of $[A' + B']^{-1}$ and the substitution $\beta_l E = \exp(-u/2)$. A Gompertz-function is defined by:

$$\left.\begin{aligned}\exp(-\xi \exp(-u/2)) &= 1 - \xi \exp(-u/2) + \frac{1}{2!}\xi^2 \exp(-u). - .. + = \\ &= 1 + \sum_{k=1}^{\infty} \frac{1}{k!}(-1)^k \xi^k \exp(-ku/2)\end{aligned}\right\} \quad (17)$$

By inserting the integration boundaries $u = 2 \cdot \ln 4m \cdot E_0/(M \cdot E_I)$, i.e., $E = E_0$ and $u \to \infty$ ($E = 0$), the integration leads to a sequence of exponential functions; the power expansion is replaced by:

$$R_{CSDA} = a_1 E_0 \cdot [1 + \sum_{k=1}^{N} (b_k - b_k \exp(-g_k \cdot E_0))] \quad (N \Rightarrow \infty) \quad (17a)$$

For therapeutic protons, the restriction to $N = 2$ provides the same accuracy (Figure 2) as formula (16); the parameters are given in Table 3 ($a_1$ is the same as in Table 2).

**Table 3:** Parameters of Formula (17a); $b_1$ and $b_2$ are dimensionless; $g_1$ and $g_2$ are given in MeV$^{-1}$.

| $b_1$ | $b_2$ | $g_1$ | $g_2$ |
|---|---|---|---|
| 15.14450027 | 29.84400076 | 0.001260021 | 0.003260031 |

In the following, we will verify that the latter formula provides some advantages with respect to the inversion $E_0 = E_0(R_{CSDA})$.

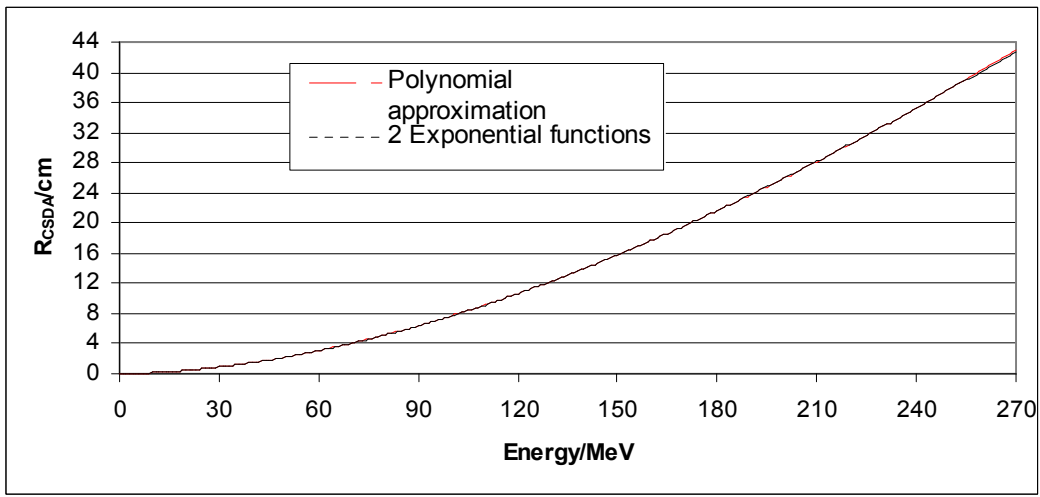

**Figure 2:** $R_{CSDA}$ calculation - comparison between a fourth-degree polynomial (equation (16)) and two exponential functions (equation (17a)).

## 2.2 The Inversion problem: calculation of $E_0(R_{CSDA})$ and $E(z)$

Above formulas can also be used for the calculation of the residual distance $R_{CSDA} - z$, relating to the residual energy $E(z)$; we have only to perform the substitutions $R_{CSDA} \rightarrow R_{CSDA} - z$ and $E_0 \rightarrow E(z)$ in these formulas. In various problems, the determination of $E_0$ or $E(z)$ as a function of $R_{CSDA}$ or $R_{CSDA} - z$ is an essential task. The power expansion implies again a corresponding series $E_0 = E_0(R_{CSDA})$ in terms of powers:

$$\left.\begin{array}{l} E_0 = \sum_{k=1}^{\infty} c_k R_{CSDA}^{\ k} \\ c_1 = 1/a_1, \quad c_2 = -a_2/a_1^3, \quad c_3 = (2a_2^2 a_1^{-1} - a_3)/a_1^4, \ldots \\ c_k = f(a_k, a_{k-1}, a_{k-2} \ldots)/a_1^{k+1} \quad (k > 3) \end{array}\right\} \quad (18)$$

The coefficients $c_k$ are calculated by a recursive procedure; we have given the first three terms in formula (18). Due to the small value of $a_1 = 6.8469 \cdot 10^{-4}$, this series is ill-posed, since there is no possibility to break off the expansion; it is divergent and the signs of the coefficients $c_k$ are alternating, see Abramowitz and Stegun (1970). The inversion procedure of equation leads to the formula:

$$\left.\begin{array}{l} E_0 = R_{csda} \sum_{i=1}^{N} c_k \exp(-\lambda_k R_{csda}) \quad (N \rightarrow \infty) \\ E(z) = (R_{csda} - z) \sum_{i=1}^{N} c_k \exp(-\lambda_k (R_{csda} - z)) \end{array}\right\} \quad (19)$$

The inverse formula of equation (17a) reads:

$$c'_k = c_k \cdot (18/10) \cdot Z \cdot \rho \cdot (75.1/E_I)^{qk} / (A_N \cdot \rho_w)$$

$$\lambda^{-1}{}'_k = \lambda^{-1}{}_k \cdot (10 \cdot \rho_w / 18) \cdot (75.1/E_I)^{pk} \cdot A_N / (\rho \cdot Z) \quad (20)$$

$$E(z) = (R_{CSDA} - z) \cdot \sum_{k=1}^{5} c'_k \cdot \exp[-(R_{CSDA} - z) \cdot \lambda'_k]$$

For therapeutic protons, a very high precision is obtained by the restriction to N = 5 (Table 4 and Figure 4). Formula (20) is also suggested by a plot $S(R_{CSDA}) = E_0(R_{CSDA})/R_{CSDA}$ according to equation (17a). This plot is shown in Figure 3 and gives rise for an expansion of $S(R_{CSDA})$ in terms of exponential functions. This plot is obtained by an interchange of the plot $E_0$ versus $R_{CSDA}$ and a calculation according to Relation.

**Table 4:** Parameters of the inversion Formula (40) with N = 5 (dimension of $c_k$: cm/MeV, $\lambda_k$: cm$^{-1}$).

| $c_1$ | $c_2$ | $c_3$ | $c_4$ | $c_5$ | $\lambda_1^{-1}$ | $\lambda_2^{-1}$ | $\lambda_3^{-1}$ | $\lambda_4^{-1}$ | $\lambda_5^{-1}$ |
|---|---|---|---|---|---|---|---|---|---|
| 96.63872 | 25.0472 | 8.80745 | 4.19001 | 9.2732 | 0.0975 | 1.24999 | 5.7001 | 10.6501 | 106.72784 |

| $P_1$ | $P_2$ | $P_3$ | $P_4$ | $P_5$ | $q_1$ | $q_2$ | $q_3$ | $q_4$ | $q_5$ |
|---|---|---|---|---|---|---|---|---|---|
| -0.1619 | -0.0482 | -0.0778 | 0.0847 | -0.0221 | 0.4525 | 0.195 | 0.2125 | 0.06 | 0.0892 |

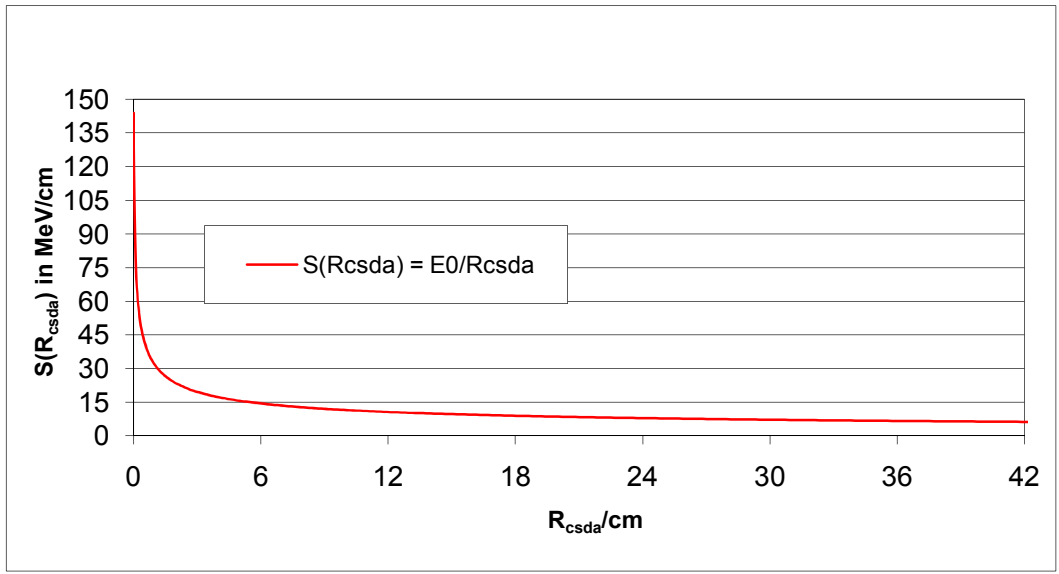

**Figure 3:** Plot $S(R_{CSDA}) = E_0/R_{CSDA}$ provides a justification of the representation of S by exponential functions.

One way to obtain the inversion Formula is to find $S(R_{CSDA})$ by a sum of exponential functions with the help of a fitting procedure. Thus it turned out that the restriction to five exponential functions is absolutely sufficient and yields a very high accuracy. A more rigorous way (mathematically) has been described in the LR of Ulmer (2007).

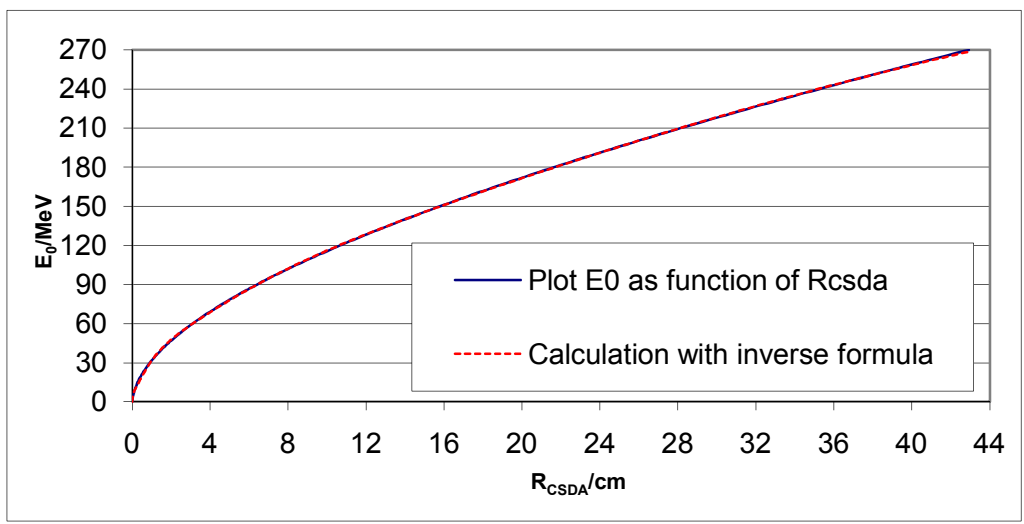

**Figure 4:** Test of the inverse Formula (40) $E_0 = E_0(R_{CSDA})$ by five exponential functions. The mean deviation amounts to 0.11 MeV. The plot results from Figure 1.

The residual energy E(z), appearing in equation (20), is the desired analytical base for all calculations of stopping power and comparisons with GEANT4. The stopping power is determined by dE(z)/dz and yields the following expression:

$$\left. \begin{array}{l} S(z) = dE(z)/dz \\ = -E(z)/(R_{CSDA} - z) + \sum_{k=1}^{N} \lambda_k E_k(z) \ (N \to \infty) \\ E_k(z) = c_k(R_{CSDA} - z) \cdot \exp[-\lambda_k(R_{CSDA} - z)] \end{array} \right\} \quad (21)$$

The aforementioned restriction to N = 5 is certainly extended to equation which can be considered as a representation of the BBE in terms of the residual energy E(z). Due to the low-energy corrections ($a_0$, $a_{shell}$, $a_{Barkas}$) the energy-transfer function dE(z)/dz remains finite for all z (i.e., $0 \leq z \leq R_{CSDA}$). This is, for instance, not true for the corresponding results according to Formulas at $z = R_{CSDA}$. The calculation of E(z) and dE/dz according to equations, referred to as LET, is presented in Figure 5. The figure shows that, within the framework of CSDA, the LET of protons is rather small, except at the distal end of the proton track.

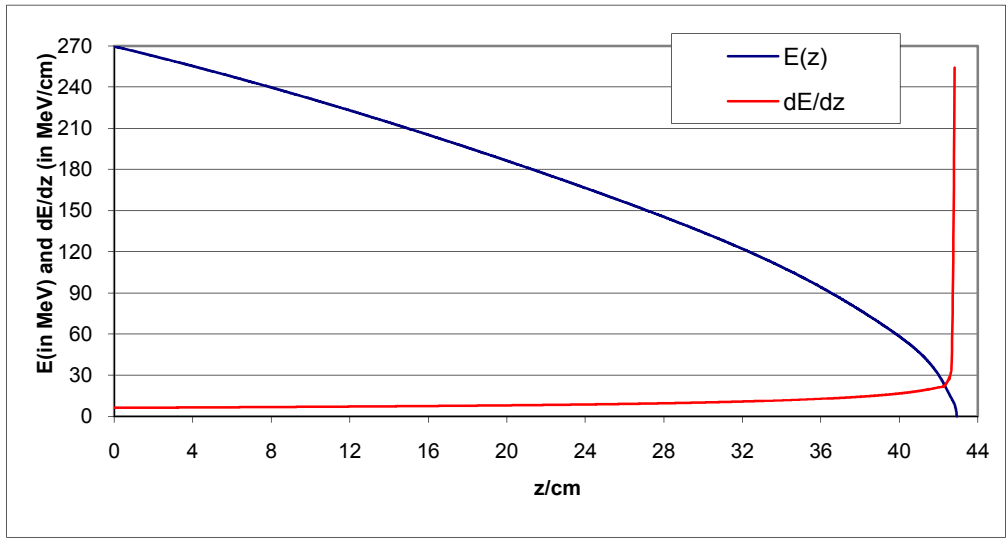

**Figure 5:** E(z) and dE(z)/dz as a function of z (LET based on CSDA); energy straggling is omitted.

A change from the interacting reference medium water to any other medium can be carried out by the calculation of $R_{CSDA}$, where the substitutions have to be performed and used in equations:

$$R_{CSDA}(medium) = R_{CSDA}(water) \cdot (Z \cdot \rho / A_N)_{water} \cdot (A_N / Z \cdot \rho)_{medium} \quad (22)$$

It is also possible to apply Formulas in a stepwise manner (e.g., voxels of CT). This procedure will not be discussed here, since it requires a correspondence between $(Z \cdot \rho / A_N)_{Medium}$ and information provided by CT, see Schneider et al (1996). With regard to heterogeneous media with only CT data as basis information the application of BBE is a more difficult task.

## 2.3 Qualitative properties of the electron transfer described by BBE and electron capture

According to BBE the energy spectrum of produced by carbon ions should be the same as that produced by protons, and the only difference between protons and carbon ions should be the intensity of the released collision electrons, i.e. the amplification factor should be 36 for carbon ions. It is well-known that this property is not valid for the following reasons: The average ionization energy for carbon ions turned out to be $E_I$ = 80 eV instead of $E_I$ = 75 eV for protons (ICRU49, Paul 2007). The paper of Paul (2007) is based on investigations of some other authors (Bichsel et al 2000, Dingfelder et al 1998, Kraft 200, Krämer et al 2000, Sigmund 1997)The second reason is the electron capture of the carbon ion. Thus a carbon ion can capture a free electron, which has been excited immediately before. Figure 7 shows this effect. However, only electrons with a slow relative velocity to the carbon ion can account for this process ($v_{relative}$ about 0). Since the transition time of the capture electron to a lower atomic state of the carbon ion is less than $10^{-10}$ sec with a simultaneous emission of light (UV or visible), it is possible that the captured electrons goes lost again, and only a stripping effect occurs for a short time. If the $C^{6+}$ ions has been finally transferred to a stable $C^{5+}$ ion, the identical process can be repeated until at the end track a neutral carbon atom is obtained having only a thermal energy. In the environment of the Bragg peak the effective charge of the carbon ion is about the same that of a proton, namely $+e_0$. Since the electron capture can only occur for electrons of which the relative velocity is slow, the upper energy limit of the energy exchange $E_{ex}$ is the Fermi edge $E_F$, which is for an electron gas not higher than the thermal energy $k_B T$. If the charge of carbon ion amounts to $+6 \cdot e_0$ and, at least, $> +e_0$, the environmental atomic electrons suffer lowering of the energy levels due to the Coulomb interaction, which leads to an increase of $E_I$. Therefore the stated value of $E_I$ = 80 eV represents an average value produced the fast carbon ion starting with $+6 \cdot e_0$ and ending with an uncharged, neutral carbon atom.

**Qualitative figure of projectile interaction of a charged particle BBE:**

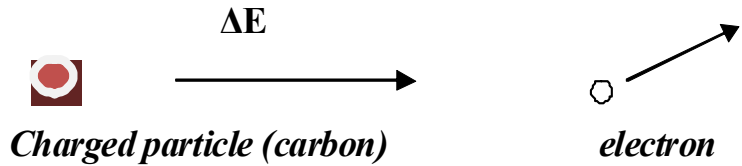

*Charged particle (carbon)*     *electron*

**Relative velocity between carbon and electron v = 0 (transition time < $10^{-10}$ sec) :**

**Electron capture by carbon ion**

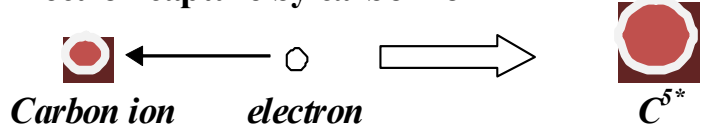

*Carbon ion    electron*                          $C^{5*}$

**Figure 7:** Excitation of an atomic electron by the collision interaction of a fast carbon ion with an atomic electron and the reversal process of the electron capture.

### 2.4 Application Fermi-Dirac statistics to electron capture

In the following it is the task to obtain a quantum-statistical description of electron capture and stripping of electrons, i.e. those electrons which reduce the effective charge of the carbon ion for a short time and go lost before a transition to a stable atomic state of carbon can occur. For this purpose we consider the quantum statistical energy exchange $E_{ex}$ between projectile particle such as proton, He ion or carbon ion. The related mathematical procedure can be used to describe processes like energy straggling, lateral scatter and energy/charge exchange between projectile ion and released electrons below the Fermi edge $E_F$. However, before we can account for the latter problem we have to consider the related mathematical tools.

In general, if H represents the Hamiltonian (either non-relativistic or relativistic) and f(H) an operator functions, then for continuous operators H the connection holds:

$$\left. \begin{array}{l} H \cdot \Psi = E \cdot \Psi \\ f(H) \cdot \Psi = f(E) \cdot \Psi \end{array} \right\} \quad (23)$$

At first we apply this relation in the non-relativistic case to derive the Gaussian convolution for the description of energy straggling. If the stopping power S(z) = dE(z)/dz of protons is calculated by BBE or by phenomenological equations (13, 22) based on classical energy dissipation, then the energy fluctuations are usually accounted for by:

$$S(z) = \int S_{Rcsda}(u) K(\sigma, u-z) du \quad (24)$$

This kernel may either be established by non-relativistic transport theory (Boltzmann equation) or, as we prefer here, by a quantum statistical derivation. Let φ be a distribution function and Φ a source function, mutually connected by the operator $F_H$ (operator notation of a canonical ensemble):

$$\left.\begin{array}{l} \varphi = \exp(-H/E_{ex})\Phi = F_H \Phi \\ F_H = \exp(-H/E_{ex}) \end{array}\right\} \quad (25)$$

An exchange Hamiltonian H couples the source field Φ (proton fluence) with an environmental field φ by $F_H$, due to the interaction with electrons:

$$\left.\begin{array}{l} H = -\frac{\hbar^2}{2m} d^2/dz^2 \\ \exp(0.25\sigma^2 d^2/dz^2)\Phi = F_H \Phi = \varphi \\ \sigma^2 = 2\hbar^2/mE_{ex} \end{array}\right\} \quad (26)$$

It must be noted that the operator equation (26) was formally introduced (Ulmer 2010) to obtain a Gaussian convolution as Green's function and to derive the inverse convolution. $F_H$ may formally be expanded in the same fashion as the usual exponential function exp(ξ); ξ may either be a real or complex number. This expansion is referred to as Lie series of an operator function. Only in the thermal limit (equilibrium), can we write $E_{ex} = k_B T$, where $k_B$ is the Boltzmann constant and T is the temperature. This equation can be solved by the spectral theorem provided by the discipline '*functional analysis*':

$$\left.\begin{array}{l} F_H \Phi = \gamma \Phi \\ \Phi_k = \exp(-ikz)/\sqrt{2\pi} \\ F_H \Phi_k = \gamma(k)\exp(ikz)/\sqrt{2\pi} = \exp(-\sigma^2 k^2/4) \cdot \exp(ikz)/\sqrt{2\pi} \\ K(\sigma, u-z) = \int \Phi_k^*(z) \Phi_k(u) \gamma(k) dk = \frac{1}{2\pi} \int \exp(-\sigma^2 k^2/4) \exp(ik(u-z)) dk \\ K(\sigma, u-z) = \frac{1}{\sigma\sqrt{\pi}} \exp(-(u-z)^2/\sigma^2) \end{array}\right\} \quad (27)$$

The kernel *K* according to equation (27) may either be established by non-relativistic transport theory (Boltzmann equation) or, as we prefer here, by a quantum statistical derivation. It is a noteworthy result (Ulmer 2007) that a quantum stochastic partition function leads to a Gaussian kernel as a Green's function, which results from a Boltzmann distribution function and a non-relativistic exchange

Hamiltonian *H*. An operator formulation of a canonical ensemble is obtained by the following way: let $\phi$ be a distribution (or output/image) function and $\Phi$ a source function, which are mutually connected by the operator. In a 3D version, linear combinations of K(σ, u – x) and the inverse kernel $K^{-1}$ are also used in scatter problems of photons (Ulmer 2010). As an example, we consider the Schrödinger equation of a free electron transferring energy from the projectile to the environment and obeying a Boltzmann distribution function f(H) =exp(-H/$E_{ex}$):

$$H = -\frac{\hbar^2}{2m}\Delta \quad (28)$$

The above relation provides:

$$\left. \begin{array}{l} \exp(-H/E_{ex}) \cdot \exp(-i\vec{k}\cdot\vec{x})/(\sqrt{2\pi})^3 = \exp(-\hbar^2\vec{k}^2/(2mE_{ex})) \cdot \exp(-i(\vec{k}\cdot\vec{x}))/(\sqrt{2\pi})^3 \\ \vec{k}^2 = k_1^2 + k_2^2 + k_3^2 \end{array} \right\} \quad (29)$$

In the case of thermal equilibrium, we can replace the exchange energy $E_{ex}$ by $k_BT$.

**2.4.1 Dirac equation and Fermi-Dirac statistics**

With regard to our task the Dirac equation to describe the particle motion is an adequate starting-point:

$$\left. \begin{array}{l} H_D = c\vec{\alpha}\vec{p} + \beta mc^2 \\ \vec{\alpha} = \begin{pmatrix} 0 & \vec{\sigma} \\ \vec{\sigma} & 0 \end{pmatrix} \quad \beta = \begin{pmatrix} 1 & 0 \\ 0 & -1 \end{pmatrix} \\ H_D^2 = c^2p^2 + m^2c^4 \end{array} \right\} \quad (30)$$

Please note that in the notation of equation (30) $\vec{\sigma}$ refers to the Pauli spin matrices (this should not be confused with the *rms*-value σ of a Gaussian distribution function). In position representation we obtain:

$$(\beta mc^2 + \frac{\hbar c}{i} \cdot \vec{\alpha} \cdot \nabla)\psi = E_D \cdot \psi \quad (31)$$

According to Feynman 1962 we can write:

$$E_D = \pm mc^2\sqrt{1 + 2 \cdot E_{Pauli}/mc^2} \quad (32)$$

$E_{Pauli}$ is the related energy value resulting from the Pauli equation.

From the view-point of the many-particle-problem Fermi-Dirac statistics is adequate mean by the notation of operator functions:

$$f_F(\hat{H}) = \frac{1}{1+\exp[(H_D - E_F)/E_{ex}]} \cdot d_s(H_D) \quad (33)$$

$E_F$ represents the energy of the Fermi edge (usually some eV) and $d_s$ the density of states of the Hamiltonian $H_D$.

$$f_F(\hat{H})^n = [\frac{1}{1+\exp[(H_D-E_F)/E_{ex}]} \cdot d_s(H_D)]^n \quad (34)$$

We iterate equation (34) n-times and obtain:

$$f_F(\hat{H})^n = [\frac{\exp[-(H_D-E_F)/2E_{ex}]}{\exp[-(H_D-E_F)/2E_{ex}]+\exp[(H_D-E_F)/2E_{ex}]} \cdot d_s(H_D)]^n \quad (35)$$

By that, the above expression assumes the shape:

$$\begin{aligned}f_F(\hat{H})^n &= [\frac{1}{2}\frac{\exp[-(H_D-E_F)/2E_{ex}]}{\cosh[(H_D-E_F)/2E_{ex}]} \cdot d_s(H_D)]^n \\ &= [\frac{1}{2}\exp[-(H_D-E_F)/2E_{ex}] \cdot \text{sech}[-(H_D-E_F)/2E_{ex}] \cdot d_s(H_D)]^n\end{aligned} \quad (36)$$

Since $1/\cosh(\xi) = \text{sech}(\xi)$ holds, we are able to carry out the following expansion, which is given by the Euler numbers $E_l$ (see e.g. Abramowitz and Stegun 1970). Convergence is only established for $\xi \leq \pi/2$. Therefore we have derived a modified expansion which provides convergence for arbitrary arguments of $\xi$ (Ulmer and Matsinos 2010):

$$\begin{aligned}\text{sech}(\xi) &= \exp(-\xi^2) \cdot \sum_{l=0}^{\infty} \alpha_{2l} \cdot \xi^{2l} \\ \alpha_{2l} &= E_{2l}/(2l)! + \sum_{l'=1}^{l}(-1)^{l'+1} \cdot \alpha_{2l-2l'}/l'!\end{aligned} \quad (37)$$

The spectral theorem of functional analysis provides:

$$\left.\begin{array}{l}\eta(k)=[mc^2\sqrt{1+\hbar^2\cdot k^2/m^2c^2}-E_F]/E_{ex}\\ \gamma(\eta(k))=[\tfrac{1}{2}(\eta\cdot E_{ex}+E_F)/mc^2]^n\cdot\exp(-n\eta/2)\cdot\operatorname{sech}(\eta/2)^n\end{array}\right\} \quad (38)$$

By performing all integrations we obtain the distribution functions in the energy space (equation (39)) and position space (equation (39a)):

$$S_E = N_f \cdot \exp(-(E_n(k)-E_{Average,n})^2/2\sigma_E(n)^2)\cdot\sum_{l=0}^{\infty} b_l(n,mc^2)\cdot(E_n(k)/2E_{ex})^l \quad (39)$$

$$K_F = N_f \cdot \sum_{l=0}^{\infty} H_l((u-z-z_{shift}(l))/\sigma_n)\cdot B_l(n,mc^2)\cdot\exp(-(u-z-z_{shift}(l))^2/2\sigma_n^2) \quad (39a)$$

According to Bohr's formalism, Bethe et al (1953), the formula for energy straggling (or fluctuation) $S_F$ is given by:

$$S_F = \frac{1}{\sqrt{\pi}\sigma_E}\exp[-(E-E_{Average})^2/\sigma_E^2] \quad (40)$$

The fluctuation parameter $\sigma_E$ can be best determined using the method of Bethe et al (1953). Furthermore we can verify the connection between $E_{Average}$ in the theory of Bohr and the Fermi edge energy $E_F$, since $E_{Average}$ results from the repeated iteration of $E_F$.

$$\left.\begin{array}{l}\Delta\sigma_E^2 = \Delta z\cdot\tfrac{1}{2}\cdot(Z/A_N)\cdot\rho\cdot f\cdot\dfrac{2mc^2}{1-\beta^2}(1-\beta^2/2) \text{ (for finite intervals } \Delta z)\\ f=0.1535\ MeVcm^2/g\\ d\sigma_E^2/dz = \tfrac{1}{2}\cdot(Z/A_N)\cdot\rho\cdot f\cdot\dfrac{2mc^2}{1-\beta^2}(1-\beta^2/2)\end{array}\right\} \quad (41)$$

$\Delta\sigma_E^2$ contains as a factor the important magnitude $E_{max}$, that is, the maximum energy transfer from the proton to an environmental electron; it is given by $E_{max} = 2mv^2/(1-\beta^2)$. In a non-relativistic approach, we get $E_{max} = 2mv^2$. $E_{max}$ can be represented in terms of the energy E, and, for the integrations to be performed, we recall the relation $E = E(z)$ according to formula (40):

$$E_{max}(in\ keV) = \sum_{k=1}^{4} s_k \cdot E^k \quad (E\ in\ MeV) \quad (42)$$

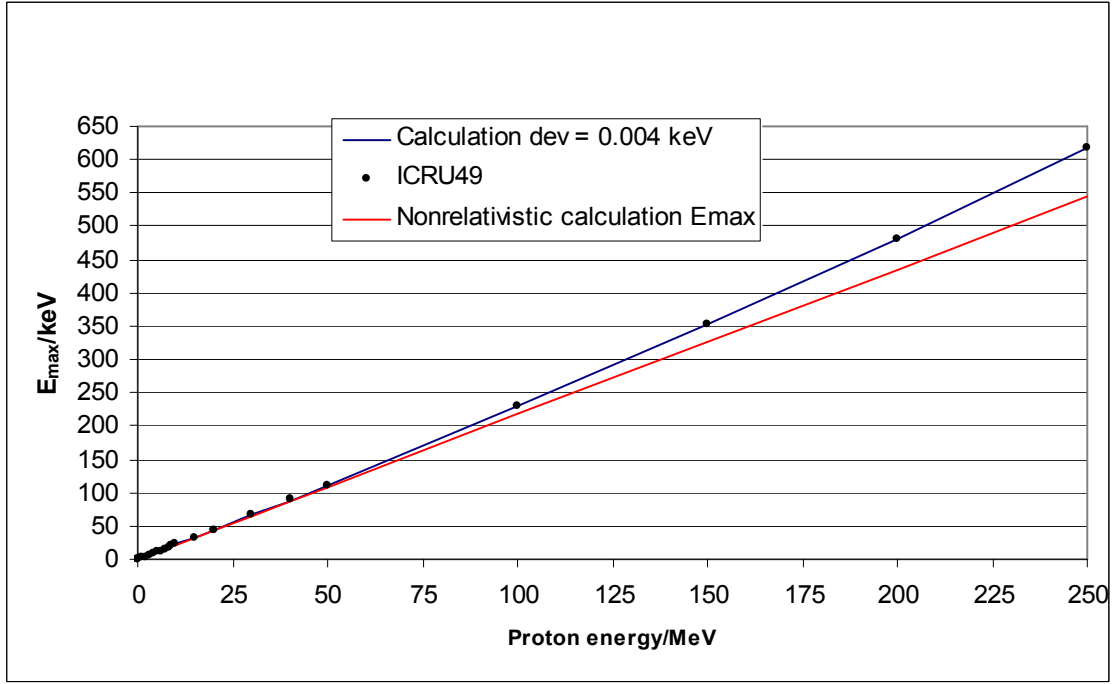

**Figure 6:** Calculation of $E_{max}$ according to equation (42). The straight line refers to the non-relativistic limit.

However, we should like to point out that according to the preceding section this determination is only valid for protons and cannot be applied to heavy ions without a change of the parameters.

**Table 5.** The parameters $s_k$ for the calculation of $E_{max}$ (formula (42)).

| $s_1$ | $s_2$ | $s_3$ | $s_4$ |
|---|---|---|---|
| 2.176519870758 | 0.001175000049 | -0.000000045000 | 0.0000000000348 |

As in a previous section we use the definition $S(z) = dE(z)/dz$ according to BBE. Since $S(z)$ is proportional to $q^2$, the following equation (44) provides $q^2(E) = q_0^2 \cdot S_E$.

$$f_F(\hat{H})\,^n S(z) = [\tfrac{1}{2}\exp[-(H_D - E_F)/2E_{ex}] \cdot \mathrm{sech}[-(H_D - E_F)/2E_{ex}] \cdot d_s(H_D)]\,^n S(z) \quad (44)$$

The transition to the integration (continuum approach up to second order) provides:

$$q^2(E) = q_0^2 [erf(E/s_E) - A \cdot (1 - E/E_0) \cdot (1 - (E/E_0) \cdot \exp(-E^2/s_E^2))]$$
$$s_E^2 = q_0^3 \cdot \pi \cdot m^2 c^4 (1 + 1/M^2 c^4) \quad (44a)$$
$$A = q_0^2 \cdot \pi^2 \cdot mc^2 / Mc^2$$

An essential result is that we are able to modify the previous formula between initial energy $E_0$ and the range $R_{csda}$:

$$Rcsda = \alpha(E_0 \cdot N / q_{eff}^2) + \beta(E_0 \cdot N / q_{eff}^2)^2 + \gamma(E_0 \cdot N / q_{eff}^2)^3 + \delta(E_0 \cdot N / q_{eff}^2)^4 \quad (45)$$
$$E_0: initial\ energy/pernucleon,\ N: nucleon\ number$$

Please note that the parameters have slightly to be modified α = 0.0069465598; β = 0.0008132157; γ = -0.00000121069; δ = 0.000000001051.

If N = 1 and $q_{eff}$ = 0.995 the above formula is valid for protons. However, it turns out that the determination of the effective charge $q_{eff}$ depends on the initial energy $E_0$. This can be verified by Figure 7.

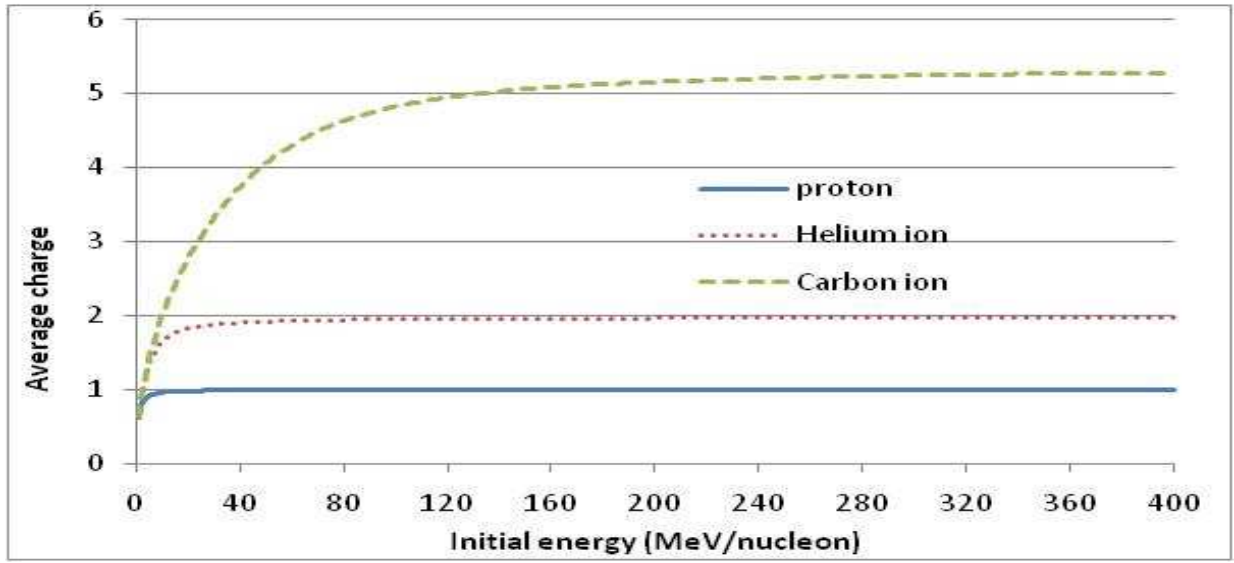

**Figure 7.** Effective charge of protons, Helium and Carbon ions as a function of the initial energy $E_0$.

### 3. Results

In the following we present results of calculations for protons, He ions and carbon ions; the initial energy amounts to 400 MeV/nucleon. This appears to be a reasonable restriction with regard to therapeutic conditions. Thus Figure 8 shows that at the end of the projectile track all charged ions nearly behave in the same manner.

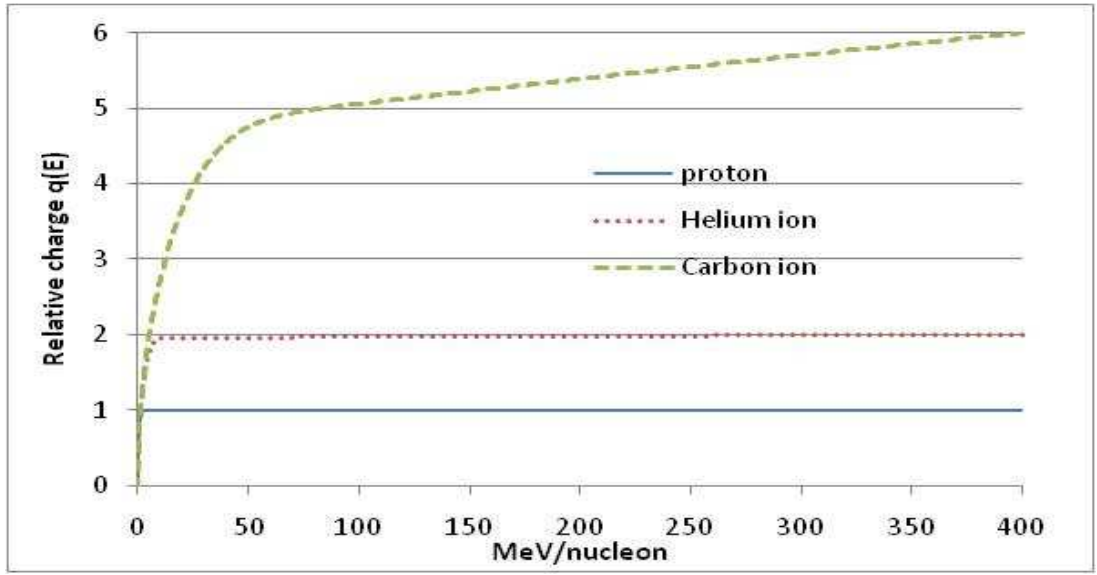

**Figure 8:** Actual charge of protons, Helium and Carbon ions in dependence of the residual energy /MeV/nucleon).

The following figure provides a more detailed behavior in the low energy domain. The residual energy per nucleon amounts to 10 MeV or smaller.

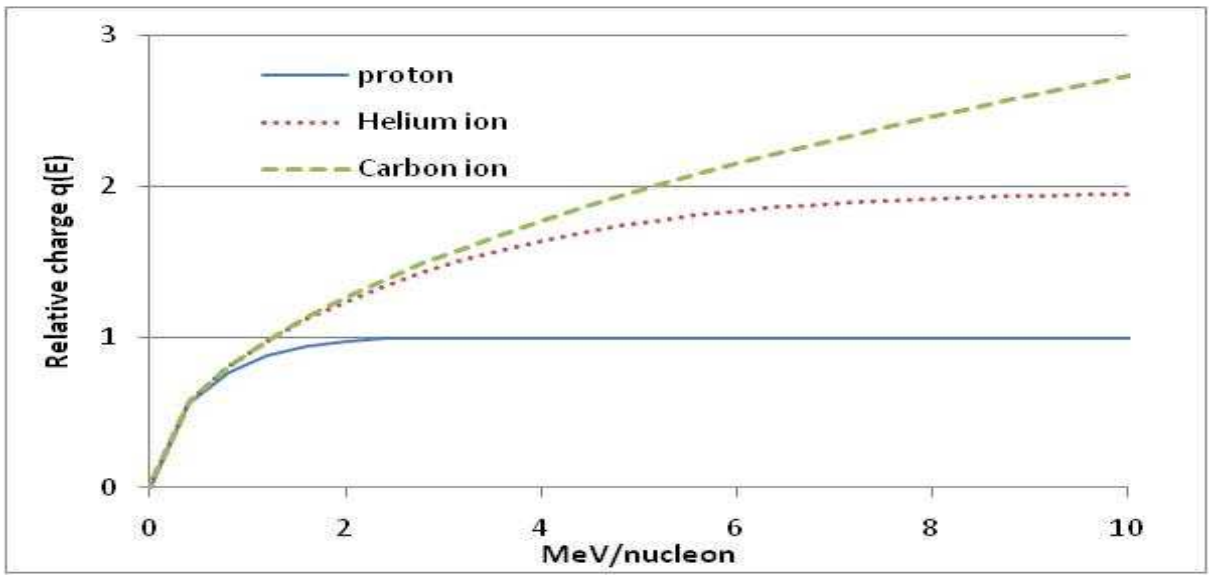

**Figure 9**: Section of the above figure for E ≤ 10 MeV.

The succeeding figure 10 presents the decrease of the actual charge of carbon ions in dependence of the initial energy $E_0$/nucleon. Thus we can conclude that for residual energies E < 50 MeV/nucleon the behavior of the carbon ions does not depend on the initial energy $E_0$.

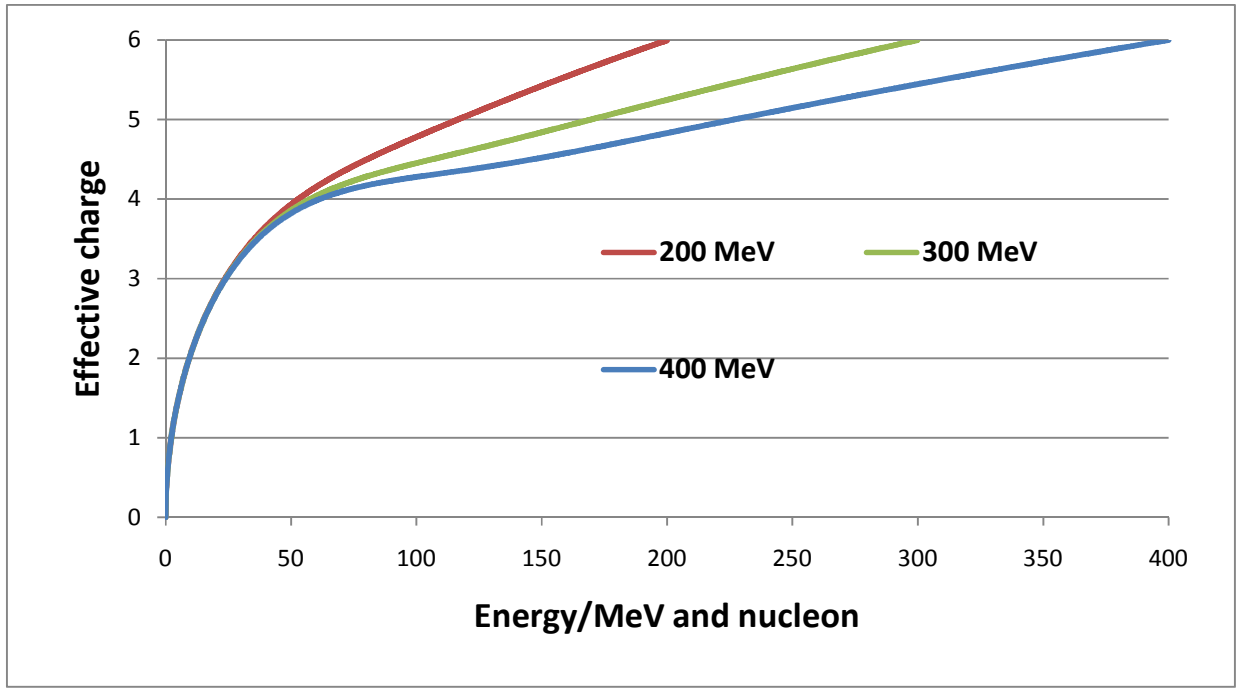

**Figure 10:** Effective charge q(E) of carbon ions in dependence of the initial energy for the cases $E_0$ = 200, 300 and 400 MeV/nucleon.

With regard to the therapeutic efficacy the behavior of the LET in the environment of the Bragg peak is very significant. For a comparison, we first regard a previous result (Ulmer and Matsinos 2010) referring to the LET of protons. According to Figure 11 the stopping power of protons at the end track depends significantly on the initial energy $E_0$ and on the beam-line (energy spectrum at the impinging plane). The electron capture of the proton at the end track is ignored. However, the previous figure 9 clearly shows that with regard to protons the electron capture only becomes more and more significant, when the actual proton energy is smaller than E = 2 MeV. The electron capture of protons at the end track would make the LET of protons zero independent of the initial energy.

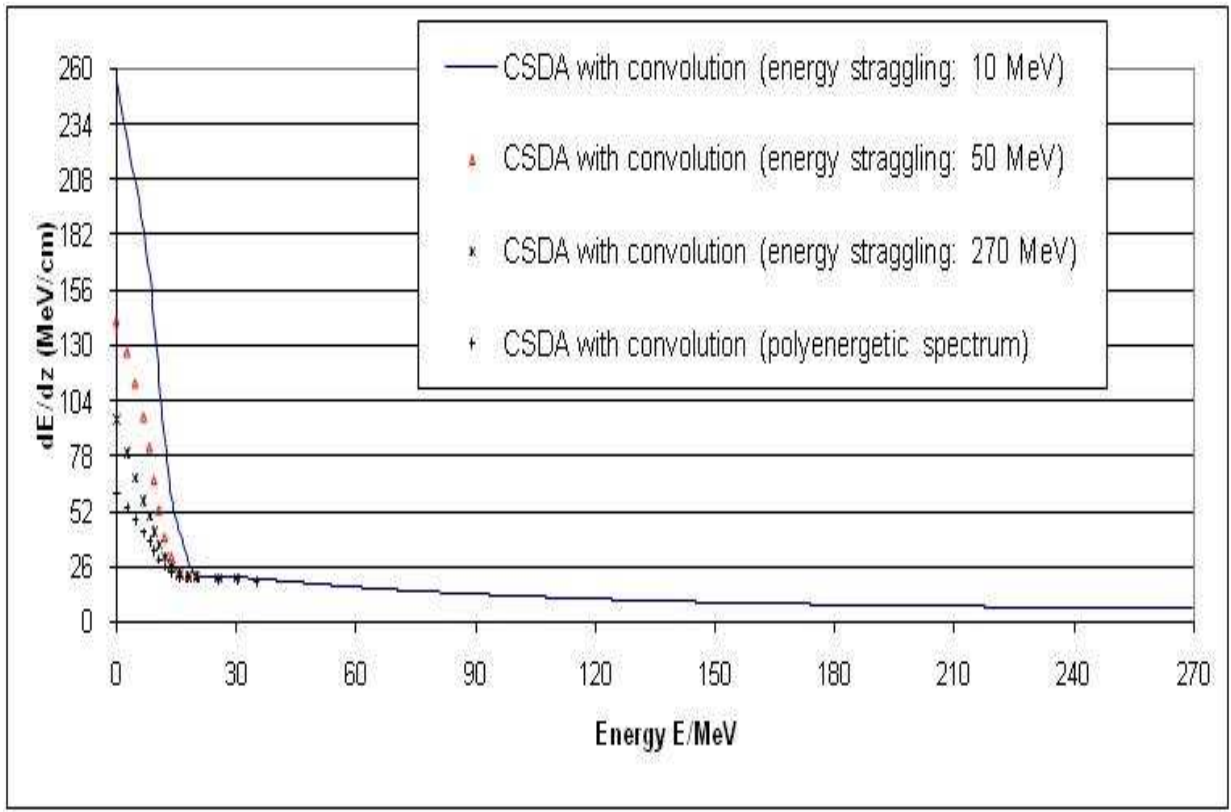

**Figure 11:** Stopping power of protons in dependence of the energy straggling (mono-energetic and polychromatic protons)

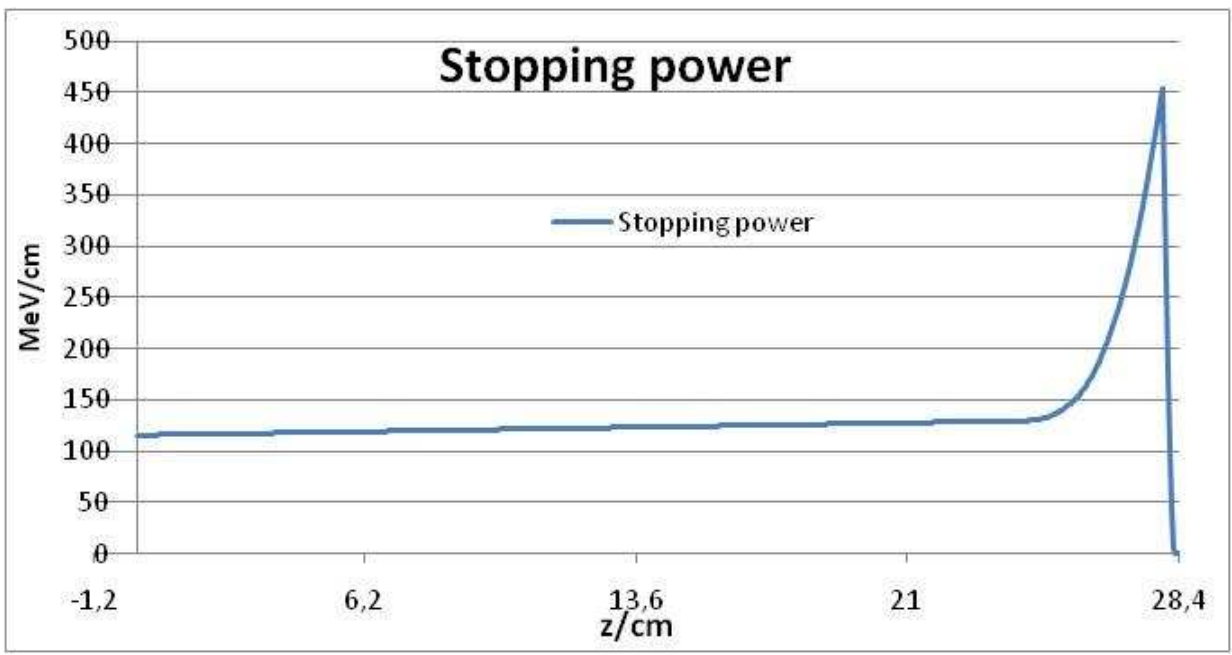

**Figure 12:** Stopping power of 400 MeV carbon ions based on the csda-approach.

The succeeding figure 13 presents E(z) and S(z) = dE(z)/dz of protons and S(z) of carbon ions with taking account for electron capture. The initial proton energy amounts to 270 MeV, whereas the initial carbon ion energy is 400 MeV/nucleon. Most significant is the height of the Bragg peak, which is resulting from the electron capture only a factor 1.7 higher than that of protons. In both cases the csda approach is assumed. Since protons are much more influenced by energy straggling and scatter, their peak height are reduced again, whereas for carbon ions scatter and energy straggling do not play a very significant role due to the mass factor 12.

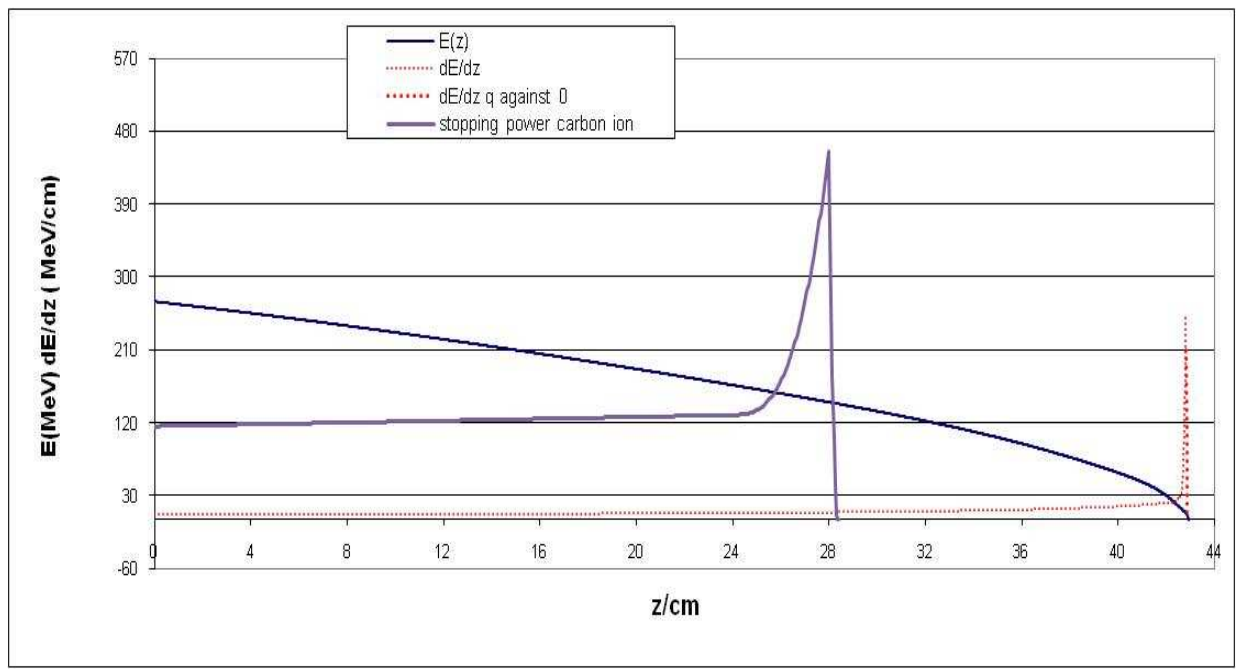

**Figure 13:** LET for mono-energetic protons and overall stopping power S(z) of carbon ions 400 MeV/nucleon.

A rigorous consideration of the LET of carbon ions is given the following figure 14. It makes only sense to consider the total energy of 4800 MeV of the carbon ions. Due to this order of magnitude E(z) of the carbon ion has not been presented in Figure 13. Energy straggling and scatter have been ignored in Figure 14, which is justified for heavy carbons. On the other side, this figure makes also apparent the well-known disadvantage of carbon ions, namely the enormous amount of energy of carbon ions in order to reach an acceptable dose distribution in the domain of the target, where a SOBP is required. With the help of GEANT4 a real depth dose curve (HIMAC, 290 MeV/nucleon) has been determined. The role of GEANT4 was only to account for the nuclear reactions, which are based in this Monte-Carlo code on an evaporation model. The electronic stopping power S(z) has been determined by the tools worked out in this communication, the electron capture effect has been accounted for. Further parameters for a calculation of S(z) have been used based on the proton calculation model (Ulmer and Matsinos 2010) by appropriate modifications. The Gaussian convolution kernels for energy straggling and lateral scatter have been rescaled according to the corresponding mass properties.

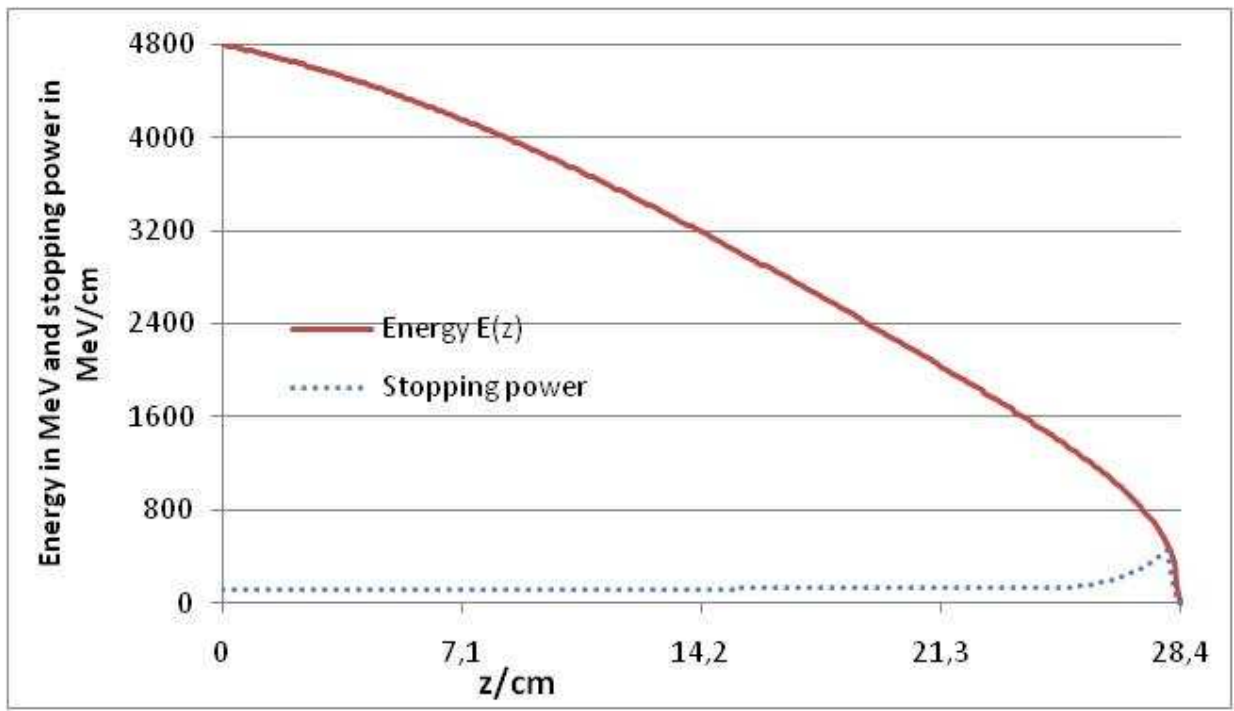

**Figure 14:** LET of carbon ions (400 MeV/nucleon).

With regard to the decrease of fluence of primary carbon ions we have derived some modifications of the corresponding decrease curves for protons. However, it appears not to be appropriate to go into further details. A further aspect is the use of the code GEANT4. Since this Monte-Carlo code represents an open programming package, some suitable additional reaction channels have been introduced.

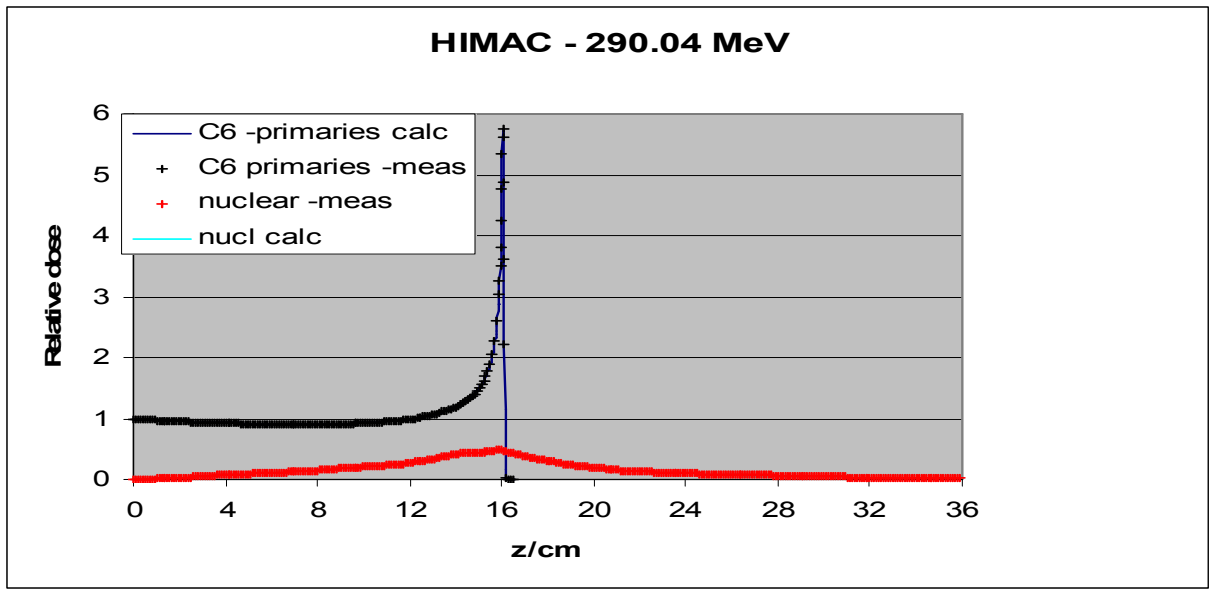

**Figure 15:** Measurement (HIMAC) and theoretical calculation of the Bragg curve of carbon ions (290 MeV/nucleon.

## 4. Discussion

The main purpose of this communication was the derivation of a systematic theory of electron capture of charged particles and the role for the LET. There are purely empirical trials to include charge capture in Monte-Carlo codes. However, it appears that a profound basis for the calculation of $q^2(E)$, $E(z)$, $S(z)$ and $R_{csda}(E_0)$ depending besides the initial energy $E_0$ also on the nuclear mass number N is required to account for further influences of Bragg curves such as the density of the medium and its nuclear mass/charge $A_N$ and Z. The unmodified use of BBE leads to wrong results and the Barkas correction, which does not affect the factor $q^2$ of BBE, only works for protons or antiprotons, whereas for projectile particles like He or carbon ions this correction cannot be considered as small. The presented theory includes the Barkas effect without any correction model.

## References

Abramowitz M and Stegun I 1970 Handbook of Mathematical Functions with Formulas, Graphs and Mathematical Tables Natural Bureau of Standards

Ashley J C, Ritchie R H, Brandt W 1974 Document No. 021195 (National Auxilliary Service, New York).

Barkas W H, Dyer N J and Heckmann H H 1963 Resolution of the Mass Anomaly Phys. Rev. Letters **11**, 26 - 29

Bethe H A 1953 Molière's theory of multiple scattering *Phys. Rev.* **89** 1256

Bethe H A and Ashkin J 1953 Passage of radiations through matter *Experimental Nuclear Physics (E. Segrè Ed.)*

Wiley New York 1953, p. 176

Betz H D 1972 Charge States and Charge-changing Cross Sections of Fast Heavy Ions Penetrating through Gaseous and Solid Media Rev. Mod. Phys. **44**, 465 – 85

Bichsel H, Hiraoka T and Omata K 2000 Aspects of Fast-Ion Dosimetry *Radiation Research* **153** 208

Bloch F 1933 Zur Bremsung rasch bewegten Teilchen beim Durchgang durch Materie *Ann der Physik* **16** 285 – 292

Boon S N 1998 Dosimetry and quality control of scanning proton beams *PhD Thesis Rijks University Groningen*

Dingfelder M, Hantke D, Inokute M and Paretzke H G 1998 *Radiation Physics and Chemistry* **53** 1 - 20

Feynman R P 1962 Quantum electrodynamics - Lecture Notes and Reprints (Benjamin, New York)

Gudowska I, Sobolevsky N, Andreo P, Belkic D and Brahme A 2004 Ion beam transport in tissue-like media using the Monte-Carlo code SHIELD-HIT *Phys. Med. Biol*. **49**, 1933 – 38

GEANT4-Documents 2005 http//geant4.web.cern.ch/geant4/G4UsersDocuments/Overview/html

Gudowska I, Sobolevsky N, Andreo P, Belkic D and Brahme A 2004 Ion beam transport in tissue-like media using the Monte-Carlo code SHIELD-HIT Phys. Med. Biol. **49**, 1933 – 38

Hollmark M, Uhrdin J, Belkic D, Gudowska I and Brahme A 2004 Influence of multiple scattering and energy loss straggling on the absorbed dose distributions of therapeutic light ion beams: 1. Analytical pencil beam model *Phys. Med. Biol.* 49 3247

Hubert F, Bimbot R and Gaurin H 1990 Range and stopping-power tables for 2.5 – 500 MeV/nucleon heavy ions in solids *Data Nucl. Data Tables* **46** 1 – 213

ICRU 1993 Stopping powers and ranges for protons and α-particles ICRU Report 49 (Bethesda, MD)

Kanai T, Kohno T, Minohara S, Sudou M, Takada E, Soga F, Kawachi K and Fukumura A 1993 Dosimetry and measured differential W values of air heavy ions *Rad. Res.* **135** 293 – 301

Kraft G 2000 Tumor therapy with heavy charged particles *Progress in Particle Nuclear Phys.* **45** 473 – 544

Kusano Y, Kanai T, Yonai S, Komori M, Ikeda N, Tachikawa Y, Ito A and Uchida H 2007 Field-size dependence of doses of therapeutic carbon beams *Med. Phys.* 34 4016

Martini A 2007 Nucleus-Nucleus Interaction Modeling and Applications in Ion Therapy Treatment Planning Doctoral dissertation at the University of Pavia

Matveev V I and Sidorov D B 2006 Effective Stopping of Fast Heavy Highly Charged Structure Ions in Collisions with Complex Atoms *JETP Letters* **84** 234 – 248

Paul H 2007 The mean ionization potential of water, and its connection to the range of energetic carbon ions in water *Nuclear Instruments and Methods in Physics Research* **B255** 435 - 437

Sigmund P 1997 Charge dependent electronic stopping power of swift non-relativistic heavy ions *Phys. Rev.* **A56** 3781 – 3793

Sigmund P and Schinner A 2002 *Nucl. Instruments and Methods* **B195** 64

Sigmund P 2006 Invited lectures presented at a symposium arranged by the Royal Danish Academy of Sciences and Letters Copenhagen, *Edited by* Peter Sigmund Matematisk-fysiske Meddelelser **52** Det Kongelige Danske Videnskabernes Selskab *The Royal Danish Academy of Sciences and Letters* Copenhagen

Sihver L, Schardt D and Kanai T 1998 Depth dose distributions of high-energy Carbon, Oxygen and Neon Beams in Water *Japanese Journal Med. Phys.* **18** 1 - 21

Ulmer W 2007 Theoretical aspects of energy range relations, stopping power and energy straggling of protons *Radiation physics and chemistry* **76** 1089 - 1107

Ulmer W 2010 Inverse problem of a linear combination of Gaussian convolution kernels (deconvolution) and some applications to proton/photon dosimetry and image processing *Inverse Problems* **26** 085002

Ulmer W and Matsinos E 2010 Theoretical methods for the calculation of Bragg curves and 3D distribution of proton beams *European Physics Journal (ST)* **190** 1 - 81

Ulmer W and Schaffner B 2011 Foundation of an analytical proton beamlet model for inclusion in a general proton dose calculation system *Radiation physics and chemistry* **80** 378 **-** 392

Yarlagadda B S, Robinson J E and Brandt W 1978 Effective-Charge Theory and the Electronic Stopping Power of Solids Phys. Rev. **B17**, 3473 – 82

Zhang R and Newhauser W D 2009 Calculation of water equivalent thickness of materials of arbitrary density, elemental composition and thickness in proton beam irradiation *Phys. Med. Biol.* **54** 1383 - 95

Ziegler J F and Manoyan J M 1988 The Stopping of Ions in Compounds Nuclear Instr. Meth. **B35**, 215 – 227